\documentclass[11pt,reqno]{amsart}
\usepackage[hidelinks]{hyperref}
\usepackage{packages, setspace, multirow}

\renewcommand{\S}{S}

\newcommand{\xpm}{x_{\alpha,\pm}}
\newcommand{\xp}{x_{\alpha,+}}
\newcommand{\xm}{x_{\alpha,-}}
\newcommand{\hol}{hol}
\newcommand{\cov}{\mathsf{cov}}

\def\XXint#1#2#3{{\setbox0=\hbox{$#1{#2#3}{\int}$}
     \vcenter{\hbox{$#2#3$}}\kern-.5\wd0}}

\theoremstyle{plain}
\newtheorem*{Thm*}{Main Theorem}

\begin{document}

  \title[Wilson Loop Area Law]{Wilson Loop Area Law for 2D Yang-Mills\\ in Generalized Axial Gauge}
  \author{Timothy Nguyen}
  \email{timothy.c.nguyen@gmail.com}
\date{\today}

\begin{abstract}
We prove that Wilson loop expectation values for arbitrary simple closed contours obey an area law up to second order in perturbative two-dimensional Yang-Mills theory. Our analysis occurs within a general family of axial-like gauges, which include and interpolate between holomorphic gauge and the Wu-Mandelstam-Liebrandt light cone gauge. Our methods make use of the homotopy-invariance properties of iterated integrals of closed one-forms, which allows us to evaluate the nontrivial integrals occurring at second order. We close with a discussion on complex gauge-fixing and deformation of integration cycles for holomorphic path integrals to shed light on some of the quantum field-theoretic underpinnings of our results.
\end{abstract}

\maketitle

\tableofcontents

\section{Introduction}

Two-dimensional Yang-Mills theory provides a rich arena for understanding quantum field theories due to the fact that the theory is soluble: exact formulas can be given for expectation values of general Wilson loop observables \cite{Mig, Dri, Wit2D, Levy}. These formulas are given by integrals of products of heat kernels on the gauge group evaluated at times equal to the areas enclosed by the Wilson loops. This suggests that Wilson loops evaluated by the methods of perturbative quantum field theory should yield a corresponding area law (heuristically, both the perturbative and exact area laws stem from the Yang-Mills path integral being formally invariant under area-preserving diffeomorphisms). Over the last twenty years, numerous checks for specialized contours to several orders \cite{BasBiaGri94, BasColNar97, StaKra} and for circular loops to all orders \cite{GioPes, GioPesRic} in perturbation theory have been made, both analytically and numerically using a variety of gauges, but no result for general contours beyond first order in perturbation theory has been established for the expected perturbative area law (which is given by a Gaussian matrix integral). Indeed, at first order, the area law is a simple consequence of Stoke's Theorem, whereas starting at second order, the nontriviality of the integrals involved in computing Wilson loop expectations have made a general analysis difficult. In this paper, we rewrite such integrals in a way that allows the homotopy-invariance properties of iterated integrals of closed one-forms to be used, thereby establishing the area law for arbitrary simple closed contours up to second order in perturbation theory. We also provide quantum field-theoretic insights as to why the area law should hold to all orders, which 
involves changing the integration cycle for a holomorphic extension of the integrand occurring in the path integral. We hope to be able to convert this analysis into rigorous mathematics, in the context of two-dimensional Yang-Mills theory, in future work.

Our main result can be described in more details as follows. Two-dimensional Yang-Mills theory takes as input a smooth surface $\Sigma$ (without boundary) equipped with an area form $d\sigma$ and a compact gauge group $G$ equipped with an ad-invariant inner product $\left<\cdot,\cdot\right>$ on its Lie algebra $\g$. The Yang-Mills action is given by
\begin{equation}
  YM(A) = \frac{1}{2\lambda_0}\int_\Sigma \left<F_A \wedge *F_A\right> \label{eq:YM}
\end{equation}
where $F_A$ is the curvature of a connection $A$ on the trivial $G$-bundle over $\Sigma$, the operator $*: \Omega^2(\Sigma) \to \Omega^0(\Sigma)$ is the Hodge star with respect to $d\sigma$, and $\lambda_0$ is a coupling constant. We consider only trivial bundles since we are interested in perturbation theory around the trivial connection.

The basic observables in Yang-Mills theory are Wilson loop observables. Given a closed oriented curve $\gm$ and a conjugation invariant function $f$ on $G$, we obtain the Wilson loop observable $$W_{f,\gm}(A) = f(\mr{hol}_\gm(A))$$
which applies $f$ to the holonomy of $A$ around $\gm$. The fundamental quantities to compute are the perturbative expectation values $\left<W_{f,\gm}\right>$ with respect to the Yang-Mills measure, which in path integral notation can be written as
\begin{equation}
  \left<W_{f,\gm}\right> \;\;``="\;\; \frac{1}{Z}\int_\A dA \, W_{f,\gm}(A)e^{-YM(A)}. \label{eq:introWLE}
\end{equation}
Here, $\frac{1}{Z}\int_\A dA$ denotes the intuition that one is supposed to integrate over the space of all connections $\A$, with $Z$ a normalizing factor. A mathematical definition of the above expectation value requires a choice of gauge-fixing and an understanding that what one obtains (in perturbation theory) is a formal series given by a Feynman diagrammatic expansion generated from the right-hand side of (\ref{eq:introWLE}). Such a definition makes no reference to integration, see Definition \ref{def:WLE}. The formal series we obtain is a series in the dimensionless coupling constant $\lambda$ defined by
\begin{equation}
\lambda = \begin{cases}
  \lambda_0|[0,1] \times [0,1]| & \Sigma = \R^2 \\
  \lambda_0|\Sigma| & \Sigma \textrm{ compact}
\end{cases}
\end{equation}
where $|U|$ denotes the area of $U \subset \Sigma$ with respect to $d\sigma$. In other words, in the Yang-Mills action, what appears is (the inverse of) an area form times $\lambda_0$ so that the separation of $\lambda_0$ from $d\sigma$ is not canonical. Thus, the dimensionless constant $\lambda$ is the meaningful perturbative parameter. 

In what follows, we work with general area forms\footnote{The work of \cite{NguSYM}, establishing various gauge-invariance properties, implies that it suffices to consider just the standard area forms on $\R^2$ and $S^2$, the main cases of interest. Nevertheless, we find it instructive to work with general area forms since the methods of \cite{NguSYM} are significantly different.} on our underlying surface $\Sigma$, where on $\R^2$ we only assume $d\sigma$ is the standard area form outside some compact set\footnote{See footnote \ref{fn:decay}.}. We can normalize $d\sigma$ so that $\Sigma$ and the unit square $[0,1]\times[0,1]$ have unit area when $\Sigma$ is compact or equal to $\R^2$, respectively. Hence, we can always assume $\lambda = \lambda_0$.

\begin{theorem}\label{mainthm} (Area Law up to Second Order)
  Fix an area form on $\R^2$, let $\gm$ be any oriented simple closed curve bounding a region $R$, and let the gauge group $G$ be any compact Lie group. Then the expectation of the Wilson loop observable $W_{f,\gm}$ with respect to generalized axial-gauge up to second order in $\lambda$ is given by a Gaussian matrix integral over $\g$:
  \begin{align}
    \left<W_{f, \gm}\right>_{gax} &= \frac{1}{(2\pi |R|\lambda)^{\dim G/2}}\int_\g e^{-|X|^2/2 |R|\lambda}f(e^X)dX + O(\lambda^3). \label{eq:main-id}
  \end{align}
  Here, $|\cdot|$ and $dX$ are the norm and volume form on $\g$ induced from its inner product.
\end{theorem}
Our terminology generalized axial-gauge refers to a general family of gauges that includes all those listed in the references used to compute Wilson loop observables. It consists of Wu-Mandelstam-Liebrandt light cone gauge,  holomorphic gauge, and all those which interpolate between the two via ``Wick rotation", see Definition \ref{Def:ax}. All these gauges on $\R^2$ are in fact equivalent up to a linear change of variables. 

Explicitly, if $G = U(N)$, $SU(N)$, or $SO(N)$ and $f = \frac{1}{N}\tr$, where $\tr$ denotes trace in the fundamental representation, then (\ref{eq:main-id}) becomes
\begin{align*}
  \left<W_{f,\gm}\right>_{gax} &= \begin{cases}
    1 - \frac{N}{2}|R|\lambda + \frac{1}{8}\left(\frac{2}{3}N^2+\frac{1}{3}\right)|R|^2\lambda^2 + O(\lambda^3) & G = U(N)\\[2ex]
  1 + \frac{C_F}{2}|R|\lambda + \frac{1}{8}\left(C_F^2-\frac{1}{6}C_AC_F\right)|R|^2\lambda^2 + O(\lambda^3) & G = SU(N),\, SO(N)
\end{cases}
\end{align*}
where $C_F$ and $C_A$ denote the Casimir numbers in the fundamental and adjoint representation, respectively.

The series generated by $\left<W_{f,\gm}\right>_{gax}$ have coefficients given by highly nontrivial integrals.  The direct evaluation of these coefficients up to second order for various types of rectangles in \cite{BasBiaGri94, BasColNar97} were nontrivial and highly involved. In contrast, we exploit homotopy-invariance properties
to compute second order coefficients for general contours. A general theory for 
the homotopy-invariance properties 
of iterated integrals dates back to the work of K. T. Chen \cite{Che}, which we became aware of through serendipitously stumbling upon the set of notes \cite{Ver}. The study of iterated integrals has found important applications in the evaluation of Feynman diagrams \cite{BogBro, BrownII}, as e.g., can be seen in the recent successes in evaluating scattering amplitudes in $N=4$ SUSY Yang-Mills theory \cite{GonSprVerVol}. 

For the special case of circular contours, using symmetry arguments, we can establish the area law to all orders in $\lambda$ on both $S^2$ on $\R^2$:

\begin{theorem}(Area Law to All Orders for Circles) Let $\gm$ be a circular contour on $S^2$ and suppose $d\sigma$ is rotationally symmetric about the center of the disk bounded by $\gm$. Let $\rho = \frac{|R||S^2\setminus R|}{|S^2|^2}$. Then
  \begin{align}
    \left<W_{f, \gm}\right>_{gax} &= \frac{1}{(2\pi \rho\lambda)^{\dim G/2}}\int_\g e^{-|X|^2/2\rho\lambda}f(e^X)dX. \label{eq:GMMS2}
  \end{align}
The same result holds on $\R^2$ by replacing $\rho$ with $|R|$.
\end{theorem}

Theorem 2 is a slight generalization of the result in \cite{GioPes}, which dealt with the standard area form on $S^2$. Although we state our results here for single Wilson loops for the sake of simplicity, they have straightforward analogs to products of Wilson loops. Theorem 1 extends to the case of when the loops are disjoint and Theorem 2 extends to the case when the loops are all circular and concentrically nested. See \cite{GioPes} for the corresponding Gaussian matrix integral formula in the latter case. The last statement about $\R^2$ should be regarded as a decompactification limit as $S^2$ increases to $\R^2$ \cite{Ngu-YM}.

The outline of this paper is as follows. We provide the basic definitions and setup in Section 2. Section 3 provides the main technical results concerning iterated integrals of Green's operators that we need to evaluate Wilson loop observables. One of our key insights involves extending known results about the homotopy invariance of iterated integrals within the class of based loops to the class of free (unbased) loops. Section 4 proves the main theorems. In the appendix, we provide an important discussion of our results from the path integral point of view. Here, we provide a formal proof of invariance of the path integral under area-preserving diffeomorphisms and we interpret our generalized axial-gauge as a (complexified) gauge-fixing procedure.

The motivation of this paper stems from the author's investigation of the conventional paradigm in physics that formal perturbation theory approximates the full quantum theory when the latter has a rigorous construction (two-dimensional Yang-Mills theory is such an example). In fact, equality to all orders in (\ref{eq:main-id}) is precisely what one would expect if perturbation theory is supposed to provide an asymptotic series for the exact Wilson loop expectation. Our work here is therefore part of a fuller discussion to be found in \cite{Ngu-YM}. See also \cite{NguSYM} for a quite different analysis of Yang-Mills theory in (bona fide) axial-gauge, in which an honest real direction is gauged away instead of a complexified one. \\

\noindent\textit{Note: } The main results of this paper are now superceded by \cite{Ngu-YM}, which establishes the area law to all orders for simple closed contours. Nevertheless, the methods here are different and moreover illuminating in that we do explicit computations of iterated integrals (\cite{Ngu-YM} uses the rather heavyweight Batalin-Vilkovisky formalism to reduce the general simple closed contour case to the circular contour case).

\section{The Setup}\label{sec:Setup}

Since we work with a trivial $G$-bundle over $\Sigma$, we can identify the space of connections $\A$ with $\Omega^1(\Sigma; \g)$, the space of $1$-forms with values in the Lie algebra $\g$ of $G$. The Yang-Mills action (\ref{eq:YM}) is invariant under the group of bundle automorphisms $\G = \mr{Maps}(\Sigma, G)$, for which a gauge-transformation $g \in \G$ acts on a connection via
\begin{equation}
  g\cdot A \mapsto gAg^{-1} + gdg^{-1}. \label{eq:gt}
\end{equation}
If we endow $\Sigma$ with a complex structure, we obtain a splitting of the complex $1$-forms $\Omega^1_c(\Sigma)$ into the bundle of complex-linear and complex anti-linear forms
$$\Omega^1_c(\Sigma) = \Omega^{1,0}(\Sigma) \oplus \Omega^{1,0}(\Sigma).$$
We can also complexify the gauge group, the Lie algebra, the space of connections, and the group of bundle automorphisms, thereby obtaining $G_c$, $\g_c$, $\A_c = \Omega^1(\Sigma; \g_c)$, and $\G_c = \mr{Maps}(\Sigma; G_c)$, respectively. The Yang-Mills action (\ref{eq:YM}) readily extends to the space of complex connections, since $\left<\cdot,\cdot\right>$ extends complex bilinearly to $\g_c$. We use a complex bilinear inner product and not a Hermitian one because then (\ref{eq:YM}) is invariant under the action of $\G_c$ (which is also given by (\ref{eq:gt})). This detail is of relevance for the quantum field theoretic interpretation of our work, which we discuss in the appendix.

\subsection{Generalized Axial Gauge}

What makes two-dimensional Yang-Mills theory tractable is that the theory on $\R^2$ becomes free in suitable gauges. Choosing global coordinates $(u,v)$ on $\R^2$ (not necessarily the standard $(x_0,x_1)$ coordinates) and writing $A = A_udu + A_vdv$, we can use a gauge transformation to eliminate the $A_v$ component. The result is a connection $A = A_udu$ which has only a single component along a coordinate direction, so that its curvature $F_A$ becomes $dA$. The vanishing of the nonlinear terms in $F_A$ makes the Yang-Mills action in this gauge purely quadratic.

This feature extends to complexified connections $A \in \Omega^1(\R^2; \g_c)$ with $u$ and $v$ a set of complex coordinates on $\bC \cong \R^2$. Thus, if $A = A_u du$ with $A_u \in \Omega^0(\R^2; \g_c)$, then the curvature $F_A$ is still $dA$. We are interested in a particular choice of complex coordinates parametrized by $\alpha \in [0,\pi/2]$, namely
$$x_{\alpha,\pm} = x_0 \pm e^{i\alpha} x_1.$$
For $\alpha = 0$, we obtain the usual light-cone coordinates $x_\pm := x_0 \pm x_1$ (with respect to the standard Minkowski metric on $\R^2$) and for $\alpha = \pi/2$ we obtain the holomorphic/antiholomorphic coordinates $z$ and $\bar z$. We can interpret intermediate $\alpha$ as being a Wick rotation between these choice of coordinates (note that the corresponding momentum variables are rotated). We need only consider $\alpha \in [0,\pi/2]$, since other values of $\alpha$ can be reduced to this case by interchanging the roles of $x_{\alpha,\pm}$ or performing the transformations $x_0 \mapsto -x_0, x_1 \mapsto -x_1$.

In fact, for $\alpha \neq 0$, since $x_{\al,\pm} = (x_0 + x_1\cos\al) \pm ix_1\sin\al$, a linear change of variables transforms $\xpm$ into $z$ and $\bar z$. Thus, if one is considering fixed $\alpha$, it suffices to work with the case of $\alpha = \pi/2$ (up to a rescaling of the area form on $\R^2$). However, it is useful to consider the limit $\alpha \to 0^+$, which is why we keep the explicit $\alpha$-dependence.

Define
$$\A_{\alpha,+} = \{A \in \A_c : A = A_+dx_{\alpha,+}\}$$
to be the set of complex connections that only have a $dx_{\al,+}$ component.

\begin{Definition}
  We say that a complex connection on $\R^2$ is in \textit{$\alpha$-gauge} if $A \in \A_{\al,+}$.
\end{Definition}

On a compact surface, axial gauges are not well-defined, since one no longer has a globally defined $\R$-valued coordinate. On the other hand, the case $\alpha = \pi/2$ does extend to compact surfaces endowed with a complex structure, since while global holomorphic coordinates do not exist, the subbundle of $(1,0)$-forms is well-defined.

\begin{Definition}
Equip $\Sigma$ with a complex structure. Then a connection $A$ on $\Sigma$ is in \textit{holomorphic gauge} if $A \in \A^{1,0} := \Omega^{1,0}(\Sigma; \g_c)$.
\end{Definition}

In particular, $\alpha$-gauge for $\alpha=\pi/2$ is holomorphic gauge on $\R^2$ with its standard complex structure.

\begin{Definition}\label{Def:ax}
  We refer to the collection of $\alpha$-gauges on $\R^2$, $0 \leq \alpha \leq \pi/2$, and holomorphic gauge for a general Riemann surface $\Sigma$ as \textit{generalized axial-gauge}. Here, we interpret the case $\alpha = 0$ as the limit $\alpha \to 0^+$, which we call \textit{Wu-Mandelstam-Liebrandt (WML) light cone gauge}.
\end{Definition}

The significance of WML light cone gauge follows from the substantial physics literature which discusses perturbative quantization in light cone gauge \cite{Lie87, BasNar96, StaKra, Bas99}, for which a suitable regularization scheme is needed in this highly singular gauge. The WML prescription is one of them, and it is prescribed by being continuously connected to holomorphic gauge via Wick rotation. The significance of $\alpha \to 0^+$ cannot be seen on the classical level, in which setting $\alpha = 0$ yields $A = A_+dx_+$ for $A$ in $0$-gauge. Rather, letting $\alpha \to 0^+$ is significant for the regularization of Feynman diagrams, see (\ref{eq:GWML}). By the remarks above, it suffices to consider holomorphic gauge and so we will primarily be focused on this case. 

The sense in which placing a connection in generalized axial-gauge is really a choice of gauge is partially addressed by the following lemma:

\begin{Lemma}\label{Lemma:CGT}
   On $\R^2$, any complex connection can be placed into generalized axial-gauge using a complex gauge transformation. On $S^2$, any complex connection not holomorphically equivalent to a non-minimal Yang-Mills connection\footnote{Such connections form an open, dense set containing a neighborhood of the trivial connection. Hence, perturbation theory around the trivial connection does not detect non-minimal Yang-Mills connections. In the higher genus case, the presence of a continuous moduli of nontrivial holomorphic structures invalidates Lemma \ref{Lemma:CGT} for a non-negligible set of connections.}
 can be placed into holomorphic gauge using a complex gauge transformation.\end{Lemma}

\Proof Without loss of generality, we consider holomorphic gauge. On $\R^2 = \bC$, the lemma is precisely the standard result that every holomorphic bundle on $\bC$ is equivalent to one with the trivial holomorphic structure. For $S^2$, this follows from the work of Atiyah and Bott \cite{AtiBot}.\End

Because Lemma \ref{Lemma:CGT} involves complex gauge transformations in $\G_c$ and not unitary (i.e. ordinary) gauge transformations in $\G$, it is not clear in what sense working with $\alpha$-gauge is really a choice of honest gauge. We provide some insightful remarks on how to interpret generalized axial-gauge in the sense most appropriate for the path integral approach to quantum field theory in the appendix, one which more adequately justifies the use of the word gauge.\\

\subsection{Propagators} Next, we analyze the kinetic operator of Yang-Mills theory in generalized axial-gauge, which allows us to define the corresponding Green's function (i.e. propagator). This allows us to define the perturbative expectation value of Wilson loop observables in terms of Wick contractions using these gauge-fixed Green's functions.

The Yang-Mills action for connections in generalized axial-gauge becomes
$$YM(A) = \frac{1}{2\lambda_0}\int_\Sigma \left<A \wedge d*dA\right>$$
since $[A,A]$ vanishes for $A \in \A_{\alpha,+}$ or $A \in \A^{1,0}$.

In holomorphic gauge and $\alpha$-gauge for $\alpha > 0$, the Yang-Mills action becomes nondegenerate in the sense that the kinetic operator is elliptic. We can see this as follows. First, we work on $\R^2$. The coordinates $\xpm$ allow a splitting of all operations into these directions. We define the complex tangent vectors
\begin{align*}
  \pd_{\xpm} &= \frac{1}{2}\left(\pd_{x_0} \pm e^{-i\alpha}\pd_{x_1}\right)
\end{align*}
which satisfy
$$\pd_{\xpm}\xpm = 1.$$
In what follows, we fix $\alpha$ and will not always indicate the explicit dependence on $\alpha$ for notational clarity.

The exterior derivative is such that $d = e_i^* \wedge \pd_{e^i}$ for any local frame $e_i$ for $T\Sigma \otimes \bC$ and corresponding $\bC$-linear dual frame $e_i^*$. In particular,
\begin{align*}
  d &= d\xp \pd_{\xp} + d\xm \pd_{\xm} \\
  & =: \pd_+ + \pd_-.
\end{align*}
We have the decomposition
$$\Omega^1 = \Omega^1_+ \oplus \Omega^1_-$$
given by those forms that have only $d\xp$ and $d\xm$ components, respectively. Since $\pd_+$ annihilates elements of $\Omega^1_+$, we have
$$YM(A) = \frac{1}{2\lambda_0}\int \left<A \wedge \pd_-*\pd_-A\right> \qquad A \in \mc{A}^\alpha_+.$$
It is easy to see that
$$D = \pd_-*\pd_-: \Omega_+^1 \to \Omega_+^1$$
is an elliptic operator for $\alpha >0$. The Green's operator for $D$ is readily constructed as follows. First we compute the Green's operator for
$$\pd_-: \Omega^0 \to \Omega^1_-.$$
This is the operator
\begin{align}
  P_{\pd_-}&: \Omega^1_- \to \Omega^0 \\
  P_{\pd_-}(x,x') &= -\frac{1}{2\pi i}\frac{dx'_+}{(x-x')_+} \in \Omega^0 \boxtimes \Omega^1_+
\end{align}
It satisfies
$$\pd_-\int_{\R^2_{x'}} P_{\pd_-}(x,x') \wedge \omega(x') = \omega(x)$$
for $\omega \in \Omega^1_-$ compactly supported. The transpose operator
\begin{align*}
P_{\pd_-}^t: \Omega^2 & \to \Omega^1_+ \\
  P_{\pd_-}^t(x,x') &= \frac{1}{2\pi i}\frac{d\xp}{(x-x')_{\alpha,+}} \in \Omega^1_+ \boxtimes \Omega^0
\end{align*}
is the Green's operator for $\pd_-: \Omega^1_+ \to \Omega^2$. Thus, we find that the Green's operator for $D$ is given by the composition
\begin{align}
  P_\alpha = P_{\pd_-}^t \circ * \circ P_{\pd_-}: \Omega^1_- \to \Omega^1_+, \qquad 0 < \alpha \leq \pi/2.  \label{eq:G}
\end{align}
and its integral kernel is an element of $\Omega^1_+ \boxtimes \Omega^1_+$. We call (\ref{eq:G}) the $\alpha$-gauge propagator. In particular, for $\alpha = \pi/2$, we have the holomorphic gauge propagator on $\R^2$
\begin{align}
  P_{hol} := P_{\pi/2} = P_{\bar\pd}^t \circ * \circ P_{\bar\pd}: \Omega^{0,1} \to \Omega^{1,0} \label{eq:GholR2}
\end{align}
with integral kernel belonging to $\Omega^{1,0} \boxtimes \Omega^{1,0}$.

In the case of the standard area form on $\R^2$, it is straightforward to obtain explicit formulas for $P_\alpha$. For the case $\alpha = \pi/2$, then with respect to holomorphic coordinates $z,w$ on $\bC$,
\begin{equation}
  P_{hol}(z,w) = \frac{1}{4\pi}dz\frac{\bar z - \bar w}{z - w}dw \label{eq:GholR2flat}
\end{equation}
which one can evaluate by computing
\begin{equation}
  P_{hol}(z,w) = \frac{dzdw}{4\pi^2}\int_\bC d\sigma(u)\frac{1}{z-u}\frac{1}{u - w} \label{eq:GholR2gen}
\end{equation}
with $d\sigma(u) = d^2u$ in (\ref{eq:GholR2}). Here, the integral (\ref{eq:GholR2gen}) must be evaluated in the sense of distributions (i.e. by evaluating the above integral as a limit of integrals over larger and larger disks\footnote{\label{fn:decay} This is where we need to place some assumptions on the behavior of $d\sigma$ at infinity so that such a limit exists. For $d\sigma$ the standard area form, for large $u$, the integrand of (\ref{eq:GholR2gen}) decays as $u^{-2} + O(u^{-3}) = |u|^{-2}e^{-2i\theta} + O(u^{-3})$ which is integrable at infinity, since the leading term vanishes when integrated over compact disks.}.) Then using Stoke's Theorem to evaluate (\ref{eq:GholR2gen}), one only picks up a contribution from integrating over small circles around $z$ and $w$, from which we obtain (\ref{eq:GholR2flat}). The same computation works for the case of general $\alpha$, from which we obtain the formula
\begin{equation}
  P_\alpha(x,x') = \frac{i}{4\pi e^{i\alpha}} dx_{\alpha,+} \frac{(x - x')_{\alpha,-}}{(x - x')_{\alpha,+}} d{x'}_{\alpha,+}, \qquad 0 < \alpha \leq \pi/2. \label{eq:Galpha}
\end{equation}
Letting $\alpha \to 0^+$ in this expression, we obtain the Wu-Mandelstam-Liebrandt light cone gauge propagator
\begin{equation}
  P_{WML}(x,x') = \lim_{\eps \to 0^+}\frac{i}{4\pi} dx_+ \frac{(x - x')_-^2}{(x - x')_+(x_-x')_- - i\eps} d{x'}_+. \label{eq:GWML}
\end{equation}
If we had set $\alpha = 0$ first in the Yang-Mills action, the kinetic operator would not be elliptic. Defining a Green's operator for the resulting operator would be ambiguous due to the need to choose a regularization scheme; the WML prescription $\alpha \to 0^+$ provides such a scheme, yielding a well-defined propagator that is continuously connected to Green's operators of elliptic operators via Wick rotation.

In the above, we worked in a coordinate-free manner as possible because then (i) it is clear that our analysis holds for general area forms on $\R^2$; (ii) we can readily adapt the above analysis to the case of holomorphic gauge on compact surfaces. For $\Sigma$ compact, the operator $\bar\pd: \Omega^\bullet(\Sigma) \to \Omega^\bullet(\Sigma)$ has a kernel and cokernel. For the simplest case $\Sigma = S^2$, $\ker \bar\pd$ consists of just constant functions. By uniformization, we can suppose that $S^2$ has the standard complex structure, otherwise we can apply a diffeomorphism (and change our area form accordingly). We then have the standard coordinate charts on the Riemann sphere $S^2 = \bC_0 \cup \bC_\infty$, related to each other by stereographic projection interchanging $z = 0$ with $\tilde z = \frac{1}{z} = \infty$. In the chart $\bC_0$, the Green's operator for $\bar\pd: \Omega^0 \to \Omega^{0,1}$ has the same expression as for the case of $\bC$:
$$P_{\bar\pd}(z,w) = -\frac{1}{2\pi i}\frac{dw}{z-w}.$$
The formula for the holomorphic gauge propagator on $S^2$ involves the appropriate modification of (\ref{eq:GholR2}), and is given by
\begin{align}
  P_{hol} = P_{\bar\pd}^t \circ \tilde* \circ P_{\bar\pd}: \Omega^{0,1} \to \Omega^{1,0} \label{eq:G2holS2}
\end{align}
with the integral kernel of $P_{hol}$ lying in $\Omega^{1,0} \boxtimes \Omega^{1,0}$ and where
\begin{align*}
  \tilde * : \Omega^0 & \to \Omega^2 \\
  f &\mapsto *\left(f - \frac{1}{|S^2|}\int_{S^2} fd\sigma\right)
\end{align*}
projects Hodge star onto the top-degree forms that integrate to zero (which are thus in the image of $\bar\pd: \Omega^{1,0} \to \Omega^2$). For concreteness, we write down an explicit formula for $P_{hol}$ when $d\sigma = \frac{4d^2z}{(1+|z|^2)^2}$ is the round area form. One can explicitly compute \cite{GioPes}
\begin{align}
  P(z,w)dzdw &= dzdw \cdot \frac{1}{4\pi^2}\int_{\bC \times \bC}  \frac{1}{z - u}\left[d\sigma(u)\delta(u -u')d^2u' - \frac{1}{|S^2|}d\sigma(u)d\sigma(u')\right]\frac{1}{u' - w} \nonumber\\
  &= dzdw\frac{1}{\pi}\frac{1}{1+|z|^2}\frac{1}{1+|w|^2}\frac{\bar z - \bar w}{z - w} \label{eq:GholS2round}
\end{align}
The above integral must be interpreted in the sense of distributions on $\bC \times \bC \subset S^2 \times S^2$ since, even though $S^2$ is compact, $dz$ and $dw$ are not globally defined coordinates. Nevertheless, we obtain the correct expression for $P(z,w)dzdw$ with respect to the holomorphic coordinates on $\bC_0 \times \bC_0$. We can replace $d\sigma$ with $d\sigma_r = \frac{4c_rd^2z}{(1+|z/r|^2)^2}$, i.e., the area form on the round sphere of radius $r$ scaled by $c_r$, and denote the corresponding propagator by $P_{\hol,S^2_r}$. Letting $r \to \infty$ and $c_r \to \frac{1}{4}$, we get
\begin{equation}
  \lim_{r \to \infty} P_{hol,S^2_r} \longrightarrow P_{hol,\R^2} \label{eq:S2toR2}
\end{equation}
i.e. the holomorphic gauge propagator on $S^2$ (\ref{eq:GholS2round}) limits to the holomorphic gauge propagator on $\R^2$ (\ref{eq:GholR2}). This decompactification limit is useful in regarding the separate cases $\R^2$ and $S^2$ in the analysis to follow as being a single case. Note however that the analysis we do holds for general area forms, not just the standard ones.

The above propagators (\ref{eq:G})--(\ref{eq:GholS2round}) all yield operators on $\g$-valued forms in the natural way. The corresponding integral kernels simply become the scalar-valued ones above tensored with the element of $\g \otimes \g$ corresponding to the identity operator on $\g$ (we identify $\g$ with $\g^*$ using the inner product on $\g$). By abuse of notation, we will use $P(x,x')$ to denote both the scalar integral kernel and $\g$-valued ones, with which one we mean clear from the context (or else we denote the latter by $P^{ab}(x,x')$, with $a,b$ denoting Lie algebra indices with respect to an orthonormal basis $e_a$ of $\g$).

\subsection{Wilson Loop Expectations}

Given $\gamma: [0,1] \to \Sigma$ a smooth closed curve, endowed with the orientation induced by its parametrization, and $f: G \to \bC$ a conjugation invariant function, we obtain the Wilson loop observable $W_{f,\gm}$ given by
$$W_{f,\gamma}(A) = f(\mr{hol}_\gm(A)),$$
where $\mr{hol}_\gm(A)$ is the holonomy of $A$ about $\gm$. Since $f$ is conjugation-invariant, $W_{f,\gamma}(A)$ is independent of the gauge-equivalence class of $A$.

Such Wilson loop observables form a dense collection of continuous gauge-invariant functions and so consistute a rich set of observables. We are interested in computing their perturbative expectation value in quantum Yang-Mills theory. Perturbation theory involves applying the Wick expansion \cite{Ngu-PI} to compute the expectation values of observables that are given as polynomials (or power series) in the basic field $A$. In generalized axial-gauge, the Yang-Mills action is free\footnote{Moreover, the Faddeev-Popov determinant is trivial.}, so that the Wick expansion is essentially formal integration against a Gaussian measure (it is formal in the sense that the propagator need not induce a positive definite pairing and the series expansions one considers need not be convergent).

We have already described the propagators we will be considering in the previous section. We thus need to describe the Wilson loop observables as a power series functional. This however is readily given by the path-ordered formal series solution of the holonomy operation. We recall this setup mostly to explain and fix notation.

Since $G$ is compact, we can assume it is embedded in a group of unitary matrices. This is convenient since then all operations involving $G$ and $\g$ can be expressed in terms of matrix multiplication within an ambient space of matrices. Parallel transport along $\gm$ involves solving the ordinary differential equation\footnote{Because $A(t)$ acts by multiplication on the left in (\ref{eq:hol}), the iterated integrals we will need to consider are composed from right to left instead of the more common left to right convention.}
\begin{align}
\begin{split}
  \dot g(t) &= -A(t)g(t)  \\
  g(0) &= 1
\end{split}\label{eq:hol}
\end{align}
where $A(t) = \dot\gm^\mu(t) \llcorner A_\mu(\gm(t))$ is a $\g$-valued function. We obtain a series solution to (\ref{eq:hol}) from the path ordered exponential:
\begin{align}
  \mr{hol}_\gm(A) &= \mc{P} \exp\left(-\int_0^1 A(t)\right) \nonumber \\
  &:= 1 + \sum_{n=1}^\infty (-1)^n\int_{1 \geq t_n \geq \ldots \geq t_1 \geq 0} A(t_n)\cdots A(t_1). \label{eq:POE}
\end{align}

Next, we can suppose $f = \tr_V \rho$, where $\rho: G \to U(V)$ is an irreducible unitary representation on a vector space $V$. Using the fact that holonomy is equivariant with respect to homomorphisms
$$\rho(\mr{hol}_\gm(A)) = \mr{hol}_\gm(\rho(A)),$$
where we also write $\rho$ for the induced morphism on Lie algebras, we have that the Wilson loop observable $W_{f,\gamma}(A)$ has a series expansion
\begin{equation}
W_{f,\gm}(A) = \tr_V(1) + \sum_{n=1}^\infty(-1)^n\int_{1 \geq t_n \geq \ldots \geq t_1 \geq 0} \tr_V(\rho(A(t_1))\cdots \rho(A(t_n)). \label{eq:WLseries}
\end{equation}
This gives the explicit representation of $W_{f,\gm}(A)$ as a power series in $A$.


The expectation value of $W_{f,\gm}(A)$ in a generalized axial-gauge involves summing over all Feynman integrals with vertices given by the terms of $W_{f,\gm}(A)$ in (\ref{eq:WLseries}) and with propagator given by the corresponding generalized propagator worked out previously. We encode the combinatorics of Feynman diagrams with the following succinct notation for the sake of mathematical completness (the initiated will already be familiar with the combinatorics while the uninitiated can consult \cite{Cos} or the appendix of \cite{Ngu-YM} for further details). However, we will only need to work to second order in this paper; the calculations in Section \ref{Sec:Proofs} make the following abstract definition more explicit.

Given the propagator $P = P^{ab}_{\mu\nu}$, which is an element of $\Omega^1(\Sigma; \g_c) \boxtimes \Omega^1(\Sigma; \g_c)$, let $\pd_P$ denote the operator which performs a Wick contraction, i.e.
$$\pd_{P} \Big(A^a_\mu(x) A^b_\nu(y)\Big) = P^{ab}_{\mu\nu}(x,y).$$
Here, we regard $A^a_\mu(x)$ as a monomial on $\A$, i.e., a complex-linear functional on $\Omega^1(X; \g_c)$ which evaluates the $a = 1,\ldots, \dim(G)$ and $\mu=0,1$ components of a connection $A$ at $x$. We extend $\pd_P$ to any polynomial $p$ of degree $n$ in $A$ by defining $\pd_P p$ to be the polynomial of degree $n-2$ obtained by contracting all possible pairs of monomials in $p$. Thus, the operation $e^{\pd_P} = 1 + \sum_{n=1}^\infty \frac{1}{n!}(\pd_P)^n$ sums over all possible numbers of Wick contractions, with the symmetry factor $1/n!$ included since the Wick contractions are unlabelled. Thus, given a polynomial $p$ in $A$, define $e^{\pd_P}p|_0$ to be the sum of all Wick contractions with no remaining uncontracted monomials, i.e., the sum of all possible Feynman diagrams with no external tails.

We can now formalize our definition of the perturbative expectation of Wilson loop observables:

\begin{Definition}\label{def:WLE}
  Let $\Sigma$ be $\R^2$ or $S^2$ equipped with an area form $d\sigma$ and pick a generalized axial-gauge. Then for any observable $O$ given by a formal power series in $A$, the expectation of $O$ in the chosen generalized axial-gauge is
  \begin{equation}
    \left<O\right>_{gax} = e^{\lambda_0\pd_{P}}O\big|_0, \label{eq:DefExp}
  \end{equation}
  where $P$ is the corresponding generalized axial-gauge propagator. We have that $\left<O\right>_{gax}$ is a formal power series in $\lambda$. In particular, for $O = W_{f,\gamma}$ or a product of such observables, we use the representation (\ref{eq:WLseries}) in (\ref{eq:DefExp}). Specializing to holomorphic gauge, we denote the corresponding expectation by $\left<O\right>_{hol}$.
\end{Definition}

Note that while $\left<O\right>_{hol}$ is well-defined for general $\Sigma$ (since we obtain a well-defined holomorphic gauge propagator $P$) the presence of zero modes in the kinetic operator $D = \bar\pd * \bar\pd: \Omega^{1,0} \to \Omega^{0,1}$ in the case of positive genus means that one cannot interpret the right-hand side of (\ref{eq:DefExp}) as a full expectation (there are residual degrees of freedom left). Hence we restrict the above definition to $\R^2$ or $S^2$.

\section{Analytic Preliminaries}

The evaluation of (\ref{eq:DefExp}) requires calculating iterated integrals of generalized axial-gauge propagators $P$ along the contour given by the Wilson loop. These integrals are highly nontrivial and do not seem, on initial inspection, to be amenable to explicit evaluation for a general contour. In this section, we develop the tools needed to evaluate such integrals.

On $\R^2$, recall that $\alpha$-gauge for fixed $\alpha > 0$ can be transformed to the case $\alpha = \pi/2$. Thus, \textit{we focus on holomorphic gauge from now on}, both on $\R^2$ and $S^2$, though our results hold for the totality of generalized axial-gauges. That is, we restrict to $P = P_{hol}$ in (\ref{eq:GholR2}) and (\ref{eq:G2holS2}).

Given that we are integrating over subsets determined by curves when evaluating Wilson loops, given a (not necessarily closed) curve $\gm$, consider the following $\delta$-current on $\gm$:
$$\delta_\gm(\omega) = \int_\gm \omega, \qquad \omega \in \Omega^1.$$
We also consider the $\delta$-current supported on $(1,0)$-forms:
$$\delta_\gm^{(0,1)}(\omega) = \int_\gm \omega^{(1,0)}.$$
Our notation reflects the fact that $\delta_\gm^{(0,1)}$ should be regarded as a singular element of $\Omega^{0,1}$, the pairing with $\omega$ being given by the wedge pairing and integration. In holomorphic gauge, we need only consider the $(0,1)$-delta current, since the corresponding propagator $P$ consists only of forms of type $(1,0)$.

\begin{Lemma}
  The function
  $$\chi_\gm(z) = \frac{1}{2\pi i}\int_\gm \frac{dw}{z-w}$$
  is smooth away from $\gm$ and satisfies $\bar\pd \chi_\gm = \delta_\gm^{(0,1)}$ in the sense of currents, i.e.
  $$\int_\bC \chi_\gm (-\bar\pd\omega) = \int_\gm \omega, \qquad \omega \in \Omega^{1,0}.$$
  Moreover, we have $\chi_\gm(z) = O(\log|z-z_0|)$ as $z$ approaches a point $z_0$ in the image of $\gm$.
\end{Lemma}

\Proof Let $f_\eps$ be a smooth compactly supported function such that $f_\eps d\bar w \to \delta_\gm^{(0,1)}$ as a current. We have
$$\chi_\eps(z) = \frac{1}{2\pi i} \int_{\bC} d\bar w dw \frac{f_\eps(w)}{z-w}$$
is a smooth function and solves $\bar\pd\chi_\eps = f_\eps d\bar w$, so that
$$\int_{\bC} \chi_\eps (-\bar\pd \omega) = \int_{\bC} f_\eps(z)d\bar w \wedge \omega, \qquad \omega \in \Omega^{1,0}.$$
Letting $\eps \to 0$, the above equations imply $\lim_{\eps \to 0}\chi_\eps = \chi_\gm$ and $\bar\pd\chi_\gm = \delta_\gm^{(0,1)}$. The last statement is clear.\End

By slight abuse of notation, we will often identify $\gm$ with its image in $\Sigma$ since $\gm$ is embedded. We also use the notational abbreviation
$$\gm^n = \overbrace{\gm \times \cdots \times \gm}^{n} \subset \Sigma^n.$$

Let $\D_0 \subset [0,1]^n$ be a domain and let 
$$\D = (\,\overbrace{\gm \times \cdots \times \gm}^{n}\,)(\D_0) \subset \gm^n$$
be its image under $\gm$ applied to each variable. Given a form $\omega \in (\Omega^1)^{\boxtimes}$, define
$$\int_{\D}\omega = \int_{\D_0}(\,\overbrace{\gm \times \cdots \times \gm}^{n}\,)^*(\omega)$$
This is just a notational shortcut so as to avoid writing pullbacks everywhere in the case where $\gm$ is not embedded. (In the end, our main results hold for $\gm$ a simple closed curve, but for the time being we can work with general curves.)

Define $\delta_{\D}$ to be the current of type $(\Omega^1)^{\boxtimes n}$ on $\Sigma^n$ which takes an element of $(\Omega^1)^{\boxtimes n}$ and integrates it over $\mc{D}$. Likewise, define $\delta_{\D}^{(0,1)}$ to be the current of type $(\Omega^{0,1})^{\boxtimes n}$ that takes an element of $(\Omega^{1,0})^{\boxtimes n}$ and integrates it along $\mc{D}$.

\begin{Corollary}\label{Cor:chi}
  For any domain $\mc{D} \subset \gm^{n}$, we have
  \begin{equation}
  \chi_\mc{D}(z_n,\ldots, z_1) = \frac{1}{(2\pi i)^n}\int_\mc{D} \frac{dw_n}{z_n-w_n}\cdots\frac{dw_1}{z_1-w_1}
  \end{equation}
  is smooth away from $\mc{D}$ and satisfies
  $$\bar\pd_{n}\cdots\bar\pd_1\chi_{\mc{D}} = \delta_{\mc{D}}^{(0,1)}$$
  in the sense of distributions.
\end{Corollary}

\Proof We approximate $\D_0$ by products of intervals and apply the previous lemma to each such product. Letting the approximation tend to $\D_0$ proves the result.\End

\noindent \textbf{Notation. } In regarding $\gm^n \subset \Sigma^n$, we will need to label the points belonging to the $i$th copy of $\Sigma$ using the subscript $i$, $1 \leq i \leq n$. However, our numbering of the copies goes from \textit{right to left} (i.e. descending order from left to right). This is so that our ordering is compatible with the path ordering in (\ref{eq:WLseries}), which involves time increasing from right to left.\\

In holomorphic gauge, the kinetic operator for Yang-Mills theory is $D = \bar\pd * \bar \pd$. Consider the operator $D_0 = *\bar\pd\pd: \Omega^0 \to \Omega^0$. A complex structure and an area form on a surface determines a metric, and it is easy to see that $D_0 = -\frac{i}{2}\Delta$, where $\Delta$ is the Laplace-Beltrami operator with respect to such a metric. Let $P_0$ be a Green's operator for $D_0$, i.e.
\begin{equation}
D_0P_0f = P_0D_0f = \begin{cases}
  f & \Sigma = \R^2 \\
  f - \bE(f) & \Sigma = S^2.
\end{cases} \label{eq:D0G0}
\end{equation}
where
$$\bE(f) = \int_{S^2} d\hat \sigma f$$
is expectation with respect to the probability measure $d\hat\sigma = \frac{1}{|\Sigma|}d\sigma$.

\begin{Lemma}\label{Lemma:GG0}
  Let $P$ denote the holomorphic gauge propagator on $\R^2$ or $S^2$. For any smooth function $f$ with compact support, we have
  \begin{align}
    P(\bar\pd f) &= \pd P_0(f).  \label{eq:id1}\\
    *\bar\pd P(\bar\pd f) &= D_0 P_0 (f). \label{eq:id2}
  \end{align}
\end{Lemma}

\Proof We easily verify
\begin{align*}
  D \pd g = \bar\pd D_0 g.
\end{align*}
for all functions $g$. Now apply $P$ to both sides and replace $g$ with $P_0(f)$ in the above to obtain (\ref{eq:id1}). Equation (\ref{eq:id2}) follows from (\ref{eq:id1}) by applying $*\bar\pd$, or else by inspection from (\ref{eq:GholR2}).\End

If $f$ is a function on $(S^2)^n$, for $1 \leq i \neq j \leq n$, define the new function $\cov_{ij}f$ on $(S^2)^{n-2}$ by
\begin{multline}
  \cov_{ij}(f) = \int_{S^2} f(x_n, \ldots, x_{i+1}, y, x_{i-1}, \ldots, x_{j+1}, y, x_{j-1}, \ldots, x_1)d\hat\sigma(y) \\
  -\int_{S^2_i \times S^2_j} f(x_n,\ldots, y_i, \ldots, y_j, \ldots, x_n)d\hat\sigma(y_i)d\hat\sigma(y_j),
\end{multline}
where $S^2_i$ and $S^2_j$ denote the copy of $S^2$ in the $i$th and $j$th entry, respectively. In other words, $\cov_{ij}$ computes the covariance in the $i$th and $j$th entries, where the covariance of two functions $f$ and $g$ is
$$\cov(f,g) = \bE(fg) - \bE(f)\bE(g).$$
Here, we interpret the space of functions on $(S^2)^n$ as (a completion of) the $n$-fold tensor product of the space of functions on $S^2$.

\begin{Lemma} \label{Lemma:IntChi}
  Let $\mc{D} \subset \gm^{2n}$ be any domain. Let $P$ be the holomorphic gauge propagator and let $\tau$ be any permutation of $\{1,\ldots, 2n\}$.
  \begin{enumerate}
    \item for $\Sigma = \R^2$, we have
 \begin{multline}
    \int_\D P(z_{\tau(1)},z_{\tau(2)})\cdots P(z_{\tau(2n-1)},z_{\tau(2n)})
    = \\
      \int_{\bC^n} \chi_\D(z_{2n},\ldots, z_1)\big|_{\{z_{\tau(1)}=z_{\tau(2)}, \;\ldots\; , \; z_{\tau(2n-1)} = z_{\tau(2n)}\}}(d\sigma)^n
    \end{multline}
where $\int_{\bC^n}$ denotes the limit obtained by integrating over an increasing sequence of polydisks.
    \item for $\Sigma = S^2$, we have
      \begin{equation}
        \int_\D \hat P(z_{\tau(1)},z_{\tau(2)})\cdots \hat P(z_{\tau(2n-1)},z_{\tau(2n)}) = \cov_{\tau(1)\tau(2)}\cdots \cov_{\tau(2n-1)\tau(2n)}(\chi_\D). \label{eq:hatG}
      \end{equation}
      where $\hat P = \frac{1}{|\Sigma|}P$.
    \end{enumerate}
\end{Lemma}

Note that these integrals of the holomorphic gauge propagator are well-defined since the latter has uniformly bounded integral kernel. (Here we assume $\gm$ is piecewise embedded so that the intersection of $\D$ with the small diagonal is negligible.)\\

\Proof Without loss of generality, we can assume $\tau$ is the identity permutation, since we can always undo this permutation by acting on the arbitrary domain $\mc{D} \subset \gm^{2n}$ by a permutation. We prove the case $n=1$, with $n > 1$ being identical.

We first consider the case $\R^2$. We have 
$$(*\pd_{\bar z_1})\pd_{\bar z_2} P(z_1,z_2) = \delta(z_1-z_2)d^2z_2$$
in the sense of distributions by Lemma \ref{Lemma:GG0}. The case $n=1$ is established if we can regulate $P$ to a smooth, compactly supported $P_\eps$ such that the following steps are valid
  \begin{align}
    \int_\D P(z_1,z_2) &= \lim_{\eps \to 0} \int_\D P_\eps(z_1,z_2) \label{eq:line1}\\
    &= \lim_{\eps \to 0}\int_{\bC^2} (\bar\pd_1\bar\pd_{2}\chi_\D)(P_\eps) \\
    &= \lim_{\eps \to 0}\int_{\bC^2} \chi_\D(\bar\pd_1\bar\pd_2P_\eps) \label{eq:limit1}\\
    &= \int_{\bC^2} \chi_\D(z_1,z_2)d\sigma(z_1)\delta(z_1-z_2)d^2z_2 \label{eq:limit2}\\
    &= \int_{\bC}\chi_\D(z,z)d\sigma(z).
  \end{align}
Here the main step that needs justification is the fourth line, since $\chi_D$ is not a smooth test function. However, this is straightforward:

Let $\psi(x)$ be a smooth bump function with $\psi \equiv 1$ on $|x| \leq 1$ and $\psi \equiv 0$ on $|x| \geq 2$. We get the corresponding long-distance and short-distance regulators:
\begin{align*}
  \tilde\psi_\eps(z) &= \psi(\eps z) \\
  \psi_\eps(z) &= 1-\psi(\eps^{-1}z).
\end{align*}
Define
$$P_\eps(z_1,z_2) = \tilde\psi_\eps(z_1)\tilde\psi_\eps(z_2)\psi_\eps(z_1 - z_2)P(z_1,z_2).$$
It is smooth and compactly supported. To justify line (\ref{eq:limit2}), we have
\begin{align}
  \lim_{\eps \to 0}\chi_D(\bar\pd_1\bar\pd_2P_\eps) &= \int \lim_{\eps \to 0}\chi_D(z_1,z_2)\tilde\psi_\eps(z_1)\tilde\psi_\eps(z_2)\bar\pd_{z_1}\bar\pd_{z_2}[\psi_\eps(z_1 - z_2) P(z_1,z_2)] + \ldots \label{eq:limit3}
\end{align}
where $\ldots$ denotes terms where at least one derivative hits the long distance cutoff function $\tilde\psi$. It is easy to see that the leading term of (\ref{eq:limit3}) is precisely (\ref{eq:limit2}) based on the local behavior of $P(z_1,z_2)$ near the diagonal.

For the remainder terms, we want to show that the $\eps \to 0$ limit vanishes. This follows readily from $\chi_D(z) = O(1/|z|)$ and $\pd_{\bar z}P = O(1/|z|)$ and that the support of $d\tilde \psi$ is on an annulus of width $O(1/\eps)$. Since
\begin{align*}
  \lim_{\eps \to 0} \int_{a/\eps}^{b/\eps} \frac{\eps}{r^2}rdr & = 0 \\
  \lim_{\eps \to 0} \int_{a/\eps}^{b/\eps} \frac{\eps^2}{r}rdr & = 0,
\end{align*}
the remainder terms $\ldots$ in (\ref{eq:limit3}) vanish as $\eps \to 0$. Note that in the above $\chi_\D$ is smooth away from $\D \subset \gm^{2n}$ and has at most logarithmic singularities. Hence, we use $(*\pd_{\bar z_1})\pd_{\bar z_2}P(z_1,z_2) \to \delta(z_1-z_2)$ away from $\D$; we can then replace integration over $\bC^n\setminus \D$ with $\bC^n$ since $\chi_{\D}$ is integrable near $\D$. Moreover, the passage from (\ref{eq:limit1}) to (\ref{eq:limit2}) means that the integral in (\ref{eq:limit2}) is obtained from integrating over an increasing sequence of polydisks, as claimed.

The case $S^2$ is analogous, where the operator $\cov$ occurs due to the $S^2$ case of (\ref{eq:D0G0}), which means
$$(*\pd_{\bar z_1})\pd_{\bar z_2} \hat P(z_1,z_2) = \frac{1}{|\Sigma|}\Big(\delta(z_1-z_2)d^2z_2 - d\hat\sigma(z_2)\Big)$$
in the sense of distributions.\End

\subsection{Iterated Integrals}

Via Lemma \ref{Lemma:IntChi}, integrals of products of the propagators $P$ along a curve can be understood in terms of integrals formed out of $\chi_{\mc{D}}$. As before, let $\gm: [0,1] \to \Sigma$ be a (not necessarily closed) curve, which we identify with its image in $\Sigma$. The domain $\mc{D} \subset \gm^n$ we are interested in is the path-ordered domain
\begin{equation}
  \gm^n_{ord} = \{(\gm(t_n), \ldots, \gm(t_1)) : 1 \geq t_n \geq \ldots \geq t_1 \geq 0\}.
\end{equation}
Indeed, this is the domain of integration that appears when one expands the path-ordered exponential defining a Wilson loop observable as in (\ref{eq:WLseries}). In fact, when $\gm$ is a closed curve, so that $\gm$ is defined on $S^1 = \R/\Z$, we will also consider the orbit of $\gm^n_{ord}$ under cyclic permutations of its entries:
\begin{align}
  \tilde \gm^n_{ord} & := \bigcup_{\tau = \textrm{cyclic permutation}} \{(\gm(t_n), \ldots, \gm(t_1)) : 1 \geq t_{\sigma(n)} \geq \ldots \geq t_{\sigma(1)} \geq 0\} \nonumber\\
  &= \{(\gm(t_n), \ldots, \gm(t_1)) : t_1+1 \geq t_n \geq t_{n-1} \geq \ldots \geq t_1,\; t_1 \in [0,1]\}
\end{align}
In other words $\gm^n_{ord}$ is isomorphic to the configuration space of $n$ ordered points on an interval whereas $\tilde \gm^n_{ord}$ is isomorphic to the configuration space of $n$ cyclically ordered points on a circle. The space $\tilde \gm^n_{ord}$ arises because the trace function occurring in (\ref{eq:WLseries}) is cyclically invariant in the Lie algebraic factors. This leads us to group together integrals over the orbits of $\gm^n_{ord}$ under cyclic permutations, i.e.  a single integral over $\tilde \gm^n_{ord}$, when we evaluate the expectation value of a Wilson loop observable.

\begin{Definition}
  Let $\gamma: [0,1] \to \Sigma$ be a curve in $\Sigma$. Given $1$-forms $\alpha_1, \ldots, \alpha_n$, define the iterated integral inductively via
  $$\int_\gamma \alpha_n \circ \cdots \circ \alpha_1 = \int_0^1 \gm^*\alpha_n(t)\left(\int_{\gm|_{[0,t]}}\alpha_{n-1} \circ \cdots \circ \alpha_1\right).$$
\end{Definition}
Thus, we have
\begin{equation}
\int_\gamma \alpha_1 \circ \cdots \circ \alpha_n = \int_{1 \geq t_n \geq \ldots \geq t_1 \geq 0} (\dot\gm \llcorner \alpha_n)(t_n)\cdots (\dot\gm\llcorner\alpha_1)(t_1).
\end{equation}
Hence, the domain of integration of an iterated integral is naturally a path ordered domain $\gm^n_{ord}$. Thus, we can write
\begin{equation}
  \chi_{\gm^n_{ord}}(z_n,\ldots,z_1) = \frac{1}{(2\pi i)^n}\int_\gm \frac{dw}{z_n - w} \circ \cdots \circ \frac{dw}{z_1 - w}.
\end{equation}

The study of iterated integrals of differential forms was pioneered by Chen \cite{Che}. The most important property for us is the homotopy invariance of such iterated integrals with respect to the endpoint preserving homotopies of $\gm$.

\begin{Lemma}\label{Lemma:Relbound}
Let $\alpha_1, \ldots, \alpha_n$ be closed $1$-forms and suppose $\alpha_{i+1} \wedge \alpha_i = 0$, $1 \leq i \leq n-1$. Then $\int_\gamma \alpha_n \circ \cdots \circ \alpha_1$ is independent of the endpoint-preserving homotopy class of $\gamma$.
\end{Lemma}

\Proof For the sake of completeness, we provide a proof which expands upon the one sketched in \cite{Ver}. Let $\gamma_0$ and $\gamma_1$ be two paths that are endpoint-preserving homotopic. Then we have a map $H: [0,1] \times [0,1] \to M$ from $\gamma_0$ to $\gamma_1$, with
\begin{align*}
  \gamma_s(t) &= H(s,t) \\
  H(s,0) &= \gamma_0(0) = \gamma_1(0)\\
  H(s,1) &= \gamma_0(1) = \gamma_1(1)
\end{align*}
for all $s$. We will show that $\tilde \alpha = H^*(\alpha_n)\left(\int_{\gamma_s|_{[0,t]}} \alpha_{n-1} \circ \cdots \circ \alpha_1\right)$ is closed. The lemma will then follow from Stoke's theorem:
\begin{align*}
  \int_{\gamma_1} \alpha_n \circ \cdots \circ \alpha_1 - \int_{\gamma_0} \alpha_n \circ \cdots \circ \alpha_1
  &= \int_{\pd ([0,1]\times [0,1])}\tilde \alpha \\
  &= \int_{[0,1]\times [0,1]} d\tilde\alpha\\
  &=0.
\end{align*}
We first start with $n=2$, since $n=1$ is precisely Stoke's Theorem. Then
\begin{align*}
  d\tilde\alpha &= d (H^*\alpha_2)\wedge\left(\int_{\gamma_s|_{[0,t]}}\alpha_1\right)  - H^*\alpha_2\wedge\left(d \int_{\gamma_s|_{[0,t]}}\alpha_1\right) \\
  &= 0 - \left[H^*\alpha_2 \wedge H^*\alpha_1 + H^*\alpha_2\wedge\left(\int_{\gamma_s|_{[0,t]}}d\alpha_1\right)\right]\\
  &= 0
\end{align*}
since $\alpha_2$ is closed and $\alpha_1\wedge\alpha_2=0$. Here we used the Fundamental Theorem of Calculus in going to the second line. We now prove $n > 2$ by induction. Namely, for $j < n$, we suppose the hypothesis $\alpha_{i+1} \wedge \alpha_i = 0$, $i = 1,\ldots, j-1$ implies that $H^*\alpha_j\left(\int_{\gamma_s|_{[0,t]}}\alpha_1 \circ \ldots \circ \alpha_{j-1}\right)$ is closed. Then
\begin{gather*}
  d\left[H^*\alpha_n\left(\int_{\gamma_s|_{[0,t]}} \alpha_{n-1} \circ \cdots \circ \alpha_1\right)\right] \hspace{4in}\\
\begin{split}
  &\hspace{.3in}= -H^*\alpha_n\wedge d\left[\int_{\gamma_s|_{[0,t]}}H^*\alpha_{n-1}(t') \left( \int_{\gamma_s|_{[0,t']}}\alpha_{n-2} \circ \cdots \circ \alpha_1\right) \right] \\
  &\hspace{.3in}= -H^*\alpha_n\wedge\left[H^*\alpha_{n-1}\left(\int_{\gamma_s|_{[0,t]}} \alpha_{n-2} \circ \cdots \circ \alpha_1\right)\right] \\
  & \hspace{.3in}\qquad - H^*\alpha_n\wedge\left[\int_{\gm_s|_{[0,t]}}d\left(H^*\alpha_{n-1}(t')\left(\int_{\gamma_{s}|_{[0,t']}}\alpha_{n-2} \circ \cdots \circ \alpha_1\right) \right)\right].
\end{split}
\end{gather*}
The hypothesis $\alpha_n \wedge \alpha_{n-1} = 0$ makes the first term vanish and the second term vanishes by the inductive hypothesis.\End

Next, we make the following important observation, which allows us to consider the invariance of iterated integrals under free homotopy classes.

\begin{Lemma}
Let $\alpha_1, \ldots, \alpha_n$ be closed $1$-forms and suppose $\alpha_{i+1} \wedge \alpha_i = 0$, $1 \leq i \leq n-1$, and $\alpha_1 \wedge \alpha_n = 0$. Let $\mc{I} \subset \Sigma$ be an open set and let $H: \mc{I} \times S^1 \to M$ be a family of loops based at points of $\mc{I}$, i.e. $\gm_x(t) = H(x,t)$ is a closed loop based at $x \in \mc{I}$. Then
  \begin{equation}
    \left(\int_{\gamma_x}\alpha_n \circ \cdots \circ \alpha_2\right)\alpha_1(x) \label{eq:form}
  \end{equation}
  defines a closed $1$-form on $\mc{I}$.
\end{Lemma}

\Proof Repeating the steps in the proof of Lemma \ref{Lemma:Relbound}, we find that
\begin{equation}
  d\int_{\gamma_x}\alpha_n \circ \ldots \circ \alpha_2 = \alpha_n(x)\left(\int_{\gm_x}\alpha_{n-1} \circ \cdots \circ \alpha_2\right) -\left(\int_{\gm_x}\alpha_n \circ \cdots \circ \alpha_3\right)\alpha_2(x) \label{eq:lowerlimit}
\end{equation}
Indeed, because now we are considering free homotopy classes, the upper and lower limits of integration in $\int_{\gamma_x}\alpha_n \circ \ldots \circ \alpha_2$ are free to vary. When we apply the Fundamental Theorem of Calculus repeatedly, we pick up the boundary terms (\ref{eq:lowerlimit}). There are no other terms in (\ref{eq:lowerlimit}) from the hypotheses $\alpha_{i+1} \wedge \alpha_{i} = 0$, $i=2,\ldots, n-1$. We now have that (\ref{eq:form}) is closed because $\alpha_2 \wedge \alpha_1 = \alpha_1 \wedge \alpha_n = 0$.\End

\begin{Lemma}
Let $\gm$ be a closed curve and let $\alpha_1, \ldots, \alpha_n$ be closed $1$-forms such that  $\alpha_{i+1} \wedge \alpha_i = 0$, $1 \leq i \leq n-1$, and $\alpha_1 \wedge \alpha_n = 0$. Then $$\delta_{\tilde \gm^n_{ord}}(\alpha_n \boxtimes \cdots \boxtimes \alpha_1) = \int_{S^1} dt_0 \int_{\gm(t_0 +\; \cdot)}\alpha_n \circ \cdots \circ \alpha_1$$
is independent of the free homotopy class of $\gm$.
\end{Lemma}

\Proof A homotopy of $\gm$ is a map $h: [0,1]_s \times S^1_t \to \Sigma$, with $h(0, t)$ being the original curve. Let $\mc{I} = [0,1]_s \times S^1_{t_0}$ and $H: \mc{I} \times S^1_t \to \Sigma$ be given by $H(s,t_0,t) = h(s,t_0+t)$. Integration over $\tilde \gm^n_{ord}$ is given by integration over $\gm(t_0 + \cdot)^{n-1}_{ord}$ followed by integration over the circle $\{s=0\} \times S^1_{t_0} \subset \mc{I}$. The first of these integrations yields precisely the closed $1$-form (\ref{eq:form}) of the previous lemma. Hence its value is unchanged as we vary $s$ in the final integral over the closed loop $\{s\} \times S^1_{t_0} \subset \mc{I}$.\End

\begin{Corollary}
  Given a closed curve $\gm$, the function $\chi_{\tilde \gm^n_{ord}}(z_n,\ldots, z_1)$ is independent of the free homotopy class of $\gm$ as long as $\gm$ avoids the points $z_1, \ldots, z_n$.
\end{Corollary}

\Proof We apply the previous lemma with $\alpha_j = \frac{dw}{z_j - w}$ which are well-defined away from the singular points $z_1,\ldots, z_n$. The required hypotheses on $\alpha_j$ are trivially satisfied.\End

Now specialize to $\gm$ a simple closed curve. In other words, $\gm$ bounds a region $R$ homeomorphic to a disk, where $R$ is uniquely determined by requiring that the orientation of $\gm$ coincides with that induced from the boundary of $R$ (in the plane, $\gm$ encircles $R$ in the counterclockwise direction). The following result is the key step which allows for explicit computation at order $\lambda^2$ in this paper:

\begin{Lemma}\label{Lemma:compChi}
  Let $\gm$ be a simple closed oriented curve on $\Sigma = \R^2$ or $S^2$ which bounds a region $R$. Then
  for all $n \geq 1$, we have
  $$\chi_{\tilde \gm^n_{ord}}(z_n,\ldots, z_1) = \begin{cases}
    \frac{(-1)^n}{(n-1)!}, \qquad z_1,\ldots, z_n \in R \\
    0 \qquad z_1,\ldots, z_n \in \Sigma\setminus\bar R.
  \end{cases}$$
  Moreover, if some proper subset of the $z_1, \ldots, z_n$ is contained in $R$, with all such points coinciding, then $\chi_{\tilde \gm^n_{ord}}(z_n,\ldots, z_1)$ vanishes.
\end{Lemma}

\Proof Suppose $z_1,\ldots, z_n \in R$ which we regard as lying inside $\R^2$ (by use of stereographic projection if $\Sigma = S^2$). By homotopy invariance, we can homotope $\gm$ to increasingly larger and larger circles $C = \{re^{i\theta}: 0 \leq \theta \leq 2\pi\}$ of radius $r$ centered at the origin. Letting the radius goes to infinity, we can conclude
\begin{align}
  \chi_{\tilde \gm^n_{ord}}(z_n,\ldots, z_1) &= \frac{(-1)^n}{(2\pi i)^n}\int_{\tilde \gm^n_{ord}}\frac{dw_n}{w_n(1-z_n/w_n)}\cdots \frac{dw_1}{w_1(1-z_1/w_1)}\nonumber\\
  & = \frac{(-1)^n}{(2\pi i)^n}\int_{\tilde C^n_{ord}}\frac{dw_n}{w_n}\cdots \frac{dw_1}{w_1}\nonumber \\
  &= \frac{(-1)^n}{(2\pi i)^n}\int_{\theta_1+2\pi \geq \theta_n \geq \cdots \geq \theta_1 \atop \theta_1 \in [0,2\pi)} id\theta_n \cdots id\theta_1 \label{eq:sym1}\\
  &= \frac{(-1)^n}{(2\pi i)^n}\frac{1}{(n-1)!}\left(\int_0^{2\pi} id\theta\right)^n \label{eq:sym2}\\
  &= \frac{(-1)^n}{(n-1)!} \nonumber
\end{align}
where in the second step, we made use of the fact that (\ref{eq:sym1}) is symmetric in the $\theta_2, \ldots, \theta_n$. If all the $z_j$ lie outside $R$, then we can simply homotope $\gm$ to a point, which shows that $\chi_{\tilde\gm^n_{ord}}(z_n,\ldots,z_1)$ vanishes.

For the second statement, consider a homotopy which shrinks $\gm$ to a small circle $C_r$ of radius $r$ surrounding the point $z \in R$ equal to the $z_1,\ldots, z_n$ that belong to $R$ (the homotopy can be made to lie inside $R$ so as to avoid the singularities $z_j$ outside $R$). If there are $m$ many of the $z_1,\ldots, z_n$ equal to $z$, $m < n$, then $\chi_{\tilde \gm^n_r}(z_n,\ldots, z_1) = \chi_{\tilde C^n_r}(z_n,\ldots, z_1)$ is of order $O(r^{n-m})$. Since $r > 0$ can be made arbitrarily small, this shows that $\chi_{\tilde \gm^n_{ord}}(z_n,\ldots, z_1)$ vanishes.\End

\section{Proof of Main Theorems}\label{Sec:Proofs}

We now have all the tools to prove our main theorems. Before we begin, we introduce some terminology to the describe the terms appearing in $\left<W_{f,\gm}\right>_{gax}$. To simplify notation, we assume $f$ is a scalar multiple of trace in the fundamental representation of $G = U(N)$, $SU(N)$, $SO(N)$ so that we only have to keep track of an orthonormal basis $e_a$ of $\g$ and not its image under a Lie algebra homomorphism $\rho$ occurring in (\ref{eq:WLseries}), though our analysis carries over straightforwardly to this more general case.

As before, we can assume our generalized axial-gauge is holomorphic gauge. The $\lambda^n$ term of $\left<W_{f,\gm}\right>_{hol}$ arises from performing all possible Wick contractions of
$$\int_{\gm^{2n}_{ord}}f(A^{a_{2n}}_w(w_{2n}) dw_{2n} \cdots A^{a_{1}}_{w}(w_{1})dw_{1})$$
using the holomorphic gauge propagator $P^{ab}_{hol}(z,w)$. In the above, the integral arises from the $2n$-order term of (\ref{eq:WLseries}), where we replace the $A$ variable standing for a general connection with the $A^{a_j}_w(w_j) dw_j$ variables that index a $\g$-valued $1$-form of type $(1,0)$ on the $j$th copy of $\Sigma$. (We can do this since we are working in holomorphic gauge.) A Wick contraction consists of choosing a pair of these variables and inserting the identity tensor into pair of $\g$-slots, and inserting the $\Omega^{1,0} \boxtimes \Omega^{1,0}$ part of the propagator into the pair of differential form slots. This factorization into a Lie algebraic contraction and analytic contractions means we can factorize the product of Wick contractions into a product of a Lie algebraic factor and an analytic factor:

\begin{Definition}\label{Def:Factor}
  An $n$th order Feynman diagram $F$ is given by a choice of grouping of $\{1,\ldots, 2n\}$ into $n$ (unordered) pairs $\{(i_1,j_1), \ldots, (i_n, j_n)\}$. The $\lambda^n$ term of $\left<W_{f,\gm}\right>_{hol}$ can be written as a sum
  $$\lambda^n\sum_F \delta_{a_{i_1}a_{j_1}}\cdots \delta_{a_{i_n}a_{j_n}}f(e_{a_{2n}}\cdots e_{a_{1}})\int_{\gm^{2n}_{ord}}\hat P_{hol}(z_{i_1}, z_{j_1})\cdots \hat P_{hol}(z_{i_n}, z_{j_n}),$$
  where $F$ ranges over all $(2n-1)!!$ Feyman diagrams of order $n$ and
  \begin{equation}
    \hat P_{hol} = \begin{cases}
      P_{hol} & \Sigma = \R^2 \\
      \frac{1}{|S^2|}P_{hol} & \Sigma = S^2.
    \end{cases}
  \end{equation}
   We refer to
  \begin{align*}
    \delta_{a_{i_1}a_{j_1}}\cdots \delta_{a_{i_n}a_{j_n}}f(e_{a_{2n}}\cdots e_{a_{1}})
  \end{align*}
  as the \textit{Lie algebraic factor} and
  \begin{align*}
     \int_{\gm^{2n}_{ord}}\hat P_{hol}(z_{i_1}, z_{j_1})\cdots \hat P_{hol}(z_{i_n}, z_{j_n})
  \end{align*}
  an the \textit{analytic factor} corresponding to the Feynman diagram $F$. In the above, repeated indices are implicitly summed over.
\end{Definition}

Observe that the Gaussian matrix integral
\begin{equation}
  \frac{1}{(2\pi |R|\lambda)^{\dim G/2}}\int_\g e^{-|X|^2/2|R|\lambda}f(e^X)dX \label{eq:GMMexp}
\end{equation}
generates a Feynman diagrammatic expansion with the same set of Lie algebraic factors as $\left<W_{f,\gm}\right>_{hol}$, since both of these expressions involve $f$ applied to an exponential. On the other hand, all the Lie algebraic factors of (\ref{eq:GMMexp}) at order $\lambda^n$ are paired with $\frac{|R|^n}{(2n)!}$. Since $f$, being proportional to $\mathrm{trace}$, is cyclically invariant, it makes sense to organize Lie algebraic factors in terms of the orbits of $F$ under the natural action of the cyclic permutation of $2n$ points. As a result, in the expansion of $\left<W_{f,\gm}\right>_{hol}$ we can replace analytic factors, which perform the path-ordered integration $\int_{\gm^{2n}_{ord}}$, with an averaged integral over the cyclic orbit of $\gm^{2n}_{ord} \subset \gm^{2n}$, i.e. with $\frac{1}{2n}\int_{\tilde\gm^{2n}_{ord}}$.

\begin{Definition}
  Given a Feynman diagram $F$, we get a corresponding \textit{cyclically-ordered analytic factor}
  \begin{align*}
     \frac{1}{2n}\int_{\tilde\gm^{2n}_{ord}}\hat P_{hol}(z_{i_1}, z_{j_1})\cdots \hat P_{hol}(z_{i_n}, z_{j_n})
  \end{align*}
\end{Definition}

Altogether, in comparing $\left<W_{f,\gamma}\right>_{hol}$ with (\ref{eq:GMMexp}), one needs to determine whether the $n$th order cyclically ordered analytic factors of $\left<W_{f,\gm}\right>_{hol}$ evaluate to $\frac{|R|^n}{(2n)!}$.\\

\noindent\textbf{Proof of Theorem 1.} We work order by order in $\lambda$. Order $n=0$ is trivial. At order $n=1$, by the above analysis, we have to show that
\begin{equation}
  \frac{1}{2}\int_{\tilde\gm^2_{ord}} P(z_1,z_2) = \frac{|R|}{2}. \label{eq:firstorder}
\end{equation}
Equation (\ref{eq:firstorder}) is easily verified using Cauchy's Theorem:
\begin{align*}
  \frac{1}{2}\int_{\tilde\gm^2_{ord}}P(z_1,z_2) &= \frac{1}{2(4\pi^2)}\int_\gm dz\int_\gm dw\int_\bC d\sigma(u)\frac{1}{z-u}\frac{1}{u - w} \\
&= -\frac{1}{2(4\pi^2)}\int_\bC d\sigma(u) \int_\gm dz\int_\gm dw \frac{1}{z-u}\frac{1}{w-u} \\
&= \frac{1}{2}\int_{R}d\sigma(u)\\
&= \frac{|R|}{2}.
\end{align*}

Order $n=2$ is the first nontrivial order. We have $3$ total Feynman diagrams, namely, $$\{(12)(34)\}, \quad \{(14)(23)\}, \quad \{(13)(24)\},$$
with the first two equivalent under cyclic permutation. From the above discussion, it suffices to show that
\begin{align}
  \int_{\gm^4_{ord}} P(z_1,z_2)P(z_3,z_4) + P(z_1,z_4)P(z_2,z_3) &= \frac{2}{4}\int_{\tilde \gm^4_{ord}} P(z_1,z_2)P(z_3,z_4) \nonumber \\
  &= \frac{2}{4!}|R|^2  \label{eq:2loop1}\\
  \int_{\gm^4_{ord}}P(z_1,z_3)P(z_2,z_4) &= \frac{1}{4}\int_{\tilde \gm^4_{ord}} P(z_1,z_3)P(z_2,z_4)\nonumber \\
  &= \frac{1}{4!}|R|^2. \label{eq:2loop2}
\end{align}
Using Lemmas \ref{Lemma:IntChi} and \ref{Lemma:compChi},
\begin{align}
  \int_{\tilde \gm^4_{ord}} P(z_1,z_2)P(z_3,z_4) &= \int_{\bC^2} \chi_{\tilde \gm^4_{ord}}(z,z,z',z')d\sigma(z)d\sigma(z') \\
  &= \frac{1}{3!}\int_{R^2} d\sigma(z)d\sigma(z')\\
  &= \frac{1}{3!}|R|^2.
\end{align}
Thus, (\ref{eq:2loop1}) holds. Similarly,
\begin{align}
  \int_{\tilde \gm^4_{ord}} P(z_1,z_3)P(z_2,z_4) &= \int_{\bC^2} \chi_{\tilde \gm^4_{ord}}(z,z',z,z')d\sigma(z)d\sigma(z') \\
  &= \frac{1}{3!}\int_{R^2}d\sigma(z)d\sigma(z')\\
  &= \frac{1}{3!}|R|^2.
\end{align}
This proves the result.\End

\begin{Remark}
  Our inability to extend Theorem 1 to $\Sigma = S^2$ lies in our inability to explicitly evaluate the expressions $\cov_{12}\cov_{34}(\chi_{\tilde \gm^4_{ord}})$ and $\cov_{13}\cov_{24}(\chi_{\tilde \gm^4_{ord}})$ occurring in Lemma \ref{Lemma:IntChi}.
\end{Remark}

To analyze perturbation theory at higher order using our methods, one has to compute integrals of $\chi_{\tilde \gm^n_{ord}}$ for $n \geq 6$. We were unable to find a general method by which to evaluate such integrals, although we can analyze the case when $\gm$ is a circle and $d\sigma$ is rotationally symmetric about the center of $\gm$.\\

\noindent\textbf{Proof of Theorem 2.} Let $F = \{(i_1 j_1), \ldots, (i_n j_n)\}$ be a Feynman diagram of order $n$ and let $\ind_R(z)$ denote the indicator function of the disk $R$ enclosed by $\gm$. Then the cyclically-ordered analytic factor corresponding to $F$ is
\begin{align*}
  \frac{1}{2n}\cov_{i_1j_1}\cdots \cov_{i_nj_n}(\chi_{\tilde \gm^{2n}_{ord}}(z_1,\ldots, z_{2n})) &=
  \frac{1}{2n(2n-1)!}\cov_{i_1j_1}\cdots \cov_{i_nj_n}(\ind_R(z_1)\cdots \ind_R(z_{2n})) \\
  &= \frac{1}{(2n)!}\Big(\cov(\ind_R(z), \ind_R(z))\Big)^n \\
  &= \frac{1}{(2n)!}\left(\frac{|R||S^2\setminus R|}{|S^2|^2}\right)^n  \label{eq:circularcase1}
\end{align*}
Here we used Lemma \ref{Lemma:IntChi} to obtain the first expression. For the right-hand side of the first equality, we use that for $z$ outside the disk $R$ bounded $\gm$, we can expand the terms $\frac{dw}{z-w} = \frac{dw}{z(1-w/z)}$ occurring in $\chi_{\tilde \gm^{2n}_{ord}}$ as a power series in $z^{-1}$. From this, we use that integrals of nonzero integer powers of $z = |z|e^{i\theta}$ about radially symmetric regions (in particular, the complement of $R$) evaluate to zero. Thus, we need only consider the integrations occurring in the $\cov$ operators to be over $R$. This yields for us the first equality. The remaining steps are straightforward computations. Consequently, every cyclically-ordered analytic factor evaluates to the one corresponding to the Gaussian matrix integral occurring in (\ref{eq:GMMS2}), thereby establishing the result.\End

The analysis of cyclically-ordered analytic factors becomes much more complicated for general contours starting at third order in perturbation theory. Indeed, at third order, we have $15$ Feynman diagrams and $5$ equivalence classes under cyclic permutation. We investigated contours which were perturbations of circles and found that cyclically-ordered analytic factors are \textit{not} proportional to $|R|^3$. Remarkably however, there are $\g$-independent linear dependences among the $5$ different Lie factors (for $\g$ any compact Lie algebra) and the cyclically-ordered analytic factors, while complicated, always conspire so that their linear combination with the Lie factors yields the area law predicted by (\ref{eq:GMMexp}). Consequently, an analysis of general contours at third order in perturbation theory and beyond, if it involves a direct analysis of Feynman diagrams, will require a refined method of understanding how the Lie algebraic and analytic factors interact.

\appendix

\section{Remarks on Path Integrals Methods and Gauge-Fixing}

Here we provide a discussion that brings to light the quantum field theoretic underpinnings of our work on Yang-Mills theory. We discuss two issues: a formal proof of an area law using the path integral and interpreting generalized axial-gauge as a gauge-fixing procedure.

The first of these goes as follows. Let $\gm$ and $\gm_0$ be two simple closed curve on $\Sigma$ such that the areas of their complementary regions agree. It it then possible to find an area-preserving diffeomorphism $\Phi: \Sigma \to \Sigma$ such that $\gamma = \Phi \circ \gm_0$ \cite{Ban}. Then we have the following chain of formal equalities:
\begin{align*}
  \left<W_{f,\gm}\right> &= \frac{1}{Z}\int_\A dA \, W_{f,\Phi \circ \gm_0}(A) e^{-YM(A)}\\
  &= \frac{1}{Z}\int_\A dA \, W_{f,\gm_0}(\Phi^*A) e^{-YM(A)}\\
  &= \frac{1}{Z}\int_\A d(\Phi^{-1})^*A \, W_{f,\gm_0}(A) e^{-YM((\Phi^{-1})^*A)}\\
  &= \frac{1}{Z}\int_\A dA \, W_{f,\gm_0}(A) e^{-YM(A)}\\
  &= \left<W_{f,\gm_0}\right>
\end{align*}
The second line is a change of variables in the computation of a Wilson loop, the third line is a formal change of variables in the path integral, and the fourth line invokes formal invariance of the ``Lebesgue measure" $dA$ under diffeomorphisms and the invariance of the Yang-Mills action under area preserving-diffeomorphisms. In this way, one expects an area law for expectation values of Wilson loop observables (in the full nonperturbative theory and in perturbation theory), since the above argument shows $\left<W_{f,\gm}\right>$ depends only on the area enclosed by $\gm$.

We now justify the use of the terminology gauge in generalized axial-gauge. Recall that perturbative quantum field theory computes quantities from the path integral by a formal application of the saddle point approximation. This generates Feynman diagrams through a procedure known as the Wick expansion. As long as one recognizes that this is an algebraic definition, and not honest integration, this procedure can be formulated as rigorous mathematics \cite{Ngu-PI}. Operations such as gauge-fixing can also be handled in this context, with the proper care.

We recall the notion of the Wick expansion, following \cite{Ngu-PI}, since we will need to make some subtle remarks concerning it. To keep matters simple, we work using explicit coordinates, though as shown in \cite{Ngu-PI}, everything we do is coordinate independent. The Wick expansion, in finite dimensions, takes as input an action functional $S(x)$ defined on a real vector space $V$ of dimension $d$ with a nondegenerate critical point $x_0 = 0$. Its output is a definition of
$$\int d^dx e^{-S(x)/\hbar}$$
as a formal series in $\hbar$, which is given by splitting $S(x) = B_{ij}x^ix^j/2 + I(x)$ as a nondegenerate quadratic part ($B_{ij}$ is a nonsingular matrix) and an interaction part $I(x)$ consisting of higher order terms in the Taylor expansion of $S(x)$ about $x_0 = 0$. One then writes
$$\int d^dx e^{-S(x)/\hbar} = \int d^dx e^{-B_{ij}x^ix^j/2\hbar}e^{-I(x)/\hbar},$$
expands $e^{-I(x)/\hbar}$ as a sum of polynomials in $x$ and $\hbar^{-1}$ and then ``integrates" each of these polynomials term by term to obtain a formal series in $\hbar$. This integration is well-defined if $B_{ij}$ is positive-definite (so that one is integrating a polynomial against a Gaussian measure); if $B_{ij}$ is merely nonsingular, we can still make sense of this procedure because we can still perform Wick contractions with the inverse matrix $B^{ij}$ \cite[Definition 1.2]{Ngu-PI}. In the simplest instance, this means we have
\begin{equation}
  \left(\frac{\det B_{ij}}{(2\pi \hbar)^d}\right)^{1/2}\int_V d^dx e^{-B_{ij}x^ix^j/2\hbar}x^ix^j = \hbar B^{ij} \label{eq:Wick}
\end{equation}
where this is equality if $B_{ij}$ is positive-definite and a definition for general $B_{ij}$ nonsingular.

If one tries to interpet the Wick expansion in generalized axial-gauge in terms of ``formal integration" as above, care is needed. This is because in say holomorphic gauge, the space $\A^{1,0} = \Omega^{1,0}(\Sigma; \g)$ is a complex vector space instead of a real one. The kinetic operator $d*d$ for Yang-Mills theory becomes $D = \bar\pd * \bar\pd$ when restricted to $\A^{1,0}$. However, the bilinear pairing $\int_\Sigma A \wedge \bar\pd * \bar\pd A$ becomes ``complex degenerate", i.e. it is degenerate regarded as a pairing on the complexification of the underlying real vector space of $\A^{1,0}$ rather than on the complex vector space $\A^{1,0}$. Indeed, this is most easily illustrated in the one-dimensional situation, in which the quadratic form $z^2$ on $\bC$ becomes complex degenerate on $\R^2$, since $z^2 = x^2 - y^2 + 2ixy$ is given by the singular matrix $\begin{pmatrix}
  1 & i \\ i & - 1
\end{pmatrix}$.
While this matrix has a trivial kernel on $\R^2$, it has nontrivial kernel on $\R^2 \otimes \bC$ and hence is not an invertible matrix. Thus, one cannot apply the Wick rule to (\ref{eq:Wick}) when $V = \R^2$ and $B_{ij}$ is the previous matrix. The real vector space underlying a complex vector must be considered when applying the Wick rule, since integration over a complex vector space only takes into account the latter's structure as a real vector space.

The way out of this predicament is as follows. Consider a finite-dimensional real vector space $V$ and then its complexification $V_c$. A (symmetric) complex bilinear form is an element of $V_c^* \otimes_\bC V_c^*$, where $V_c^*$ is the $\bC$-linear dual of $V_c$. By restriction to $V$, we obtain an $\R$-linear complex-valued bilinear form on $V$, i.e., an element of $V^* \otimes_\R V^* \otimes_\R \bC \cong V_c^* \otimes_\bC V_c^*$. We make the following trivial but key observations:

\begin{Lemma}\label{LemmaBilinear}
  Consider a nondegenerate complex bilinear form $B_c$ on $V_c$. Then it restricts to a nondegenerate complex-valued bilinear form $B$ on $V$. The dual tensor $B^{-1}$ to $B$, which is an element of $V \otimes_\R V \otimes_\R \bC$, is equal to the dual tensor $B_c^{-1} \in V_c \otimes_\bC V_c$ of $B_c$.
\end{Lemma}

\Proof Since $B_c$ is complex bilinear, if its restriction to $V$ were degenerate, it would be degenerate on $V_c$ as well. So for $B_c$ and $B$ nondegenerate, consider their dual tensors $B_c^{-1}$ and $B^{-1}$. Recall that the dual tensor $B^{-1}$ to $B$ is the (complex) pairing on $V^*$ induced from the one on $V$ via the isomorphism between $V$ and $V^*$ determined by $B$. Equivalently, $B^{-1}$ satisfies $B^{-1} \circ B = \mr{id}_V$, where $\circ$ is the natural map
$$\circ: (V \otimes V) \otimes (V^* \otimes V^*) \to V \otimes V^*$$
given by the evaluation pairing on the middle factor $V \otimes V^*$. The dual tensor $B_c^{-1}$ to $B_c$ is defined similarly. Then $B^{-1} = B_c^{-1}$ readily follows from complex-bilinearity of $B_c$.\End

\noindent\textit{Example:} Let $V = \bC^n = \R^n \otimes \bC$ and let $e_1, \ldots, e_n$ be a basis for $\R^n$, with $e_1^*,\ldots, e_n^*$ a corresponding dual basis. Consider the complex bilinear form
$$B_c = \Sigma_{i=1}^n \al_i e_i^* \otimes e_i^*$$
where $\al_i \in \bC \setminus \{0\}$ and the $e_i^*$ extend complex-linearly to $\bC$. Then $B = B_c |_{\R^n}$ and $B^{-1} = B^{-1}_c|_{\R^n}$, where
$$B_c^{-1} = \Sigma_{i=1}^n \al_i^{-1} e_i \otimes e_i.$$

The significance of Lemma \ref{LemmaBilinear} is as follows. We want to apply the Wick rule using the propagator $B_c^{-1}$. The above considerations show that this cannot be interpreted as formal integration on $V_c$, since then $B_c$ is degenerate on $(V_c)_c$, i.e. we regard $V_c$ as a real vector space and consider $B_c$ on its complexification. Instead, we can consider (any) maximal totally real subspace $V$ of $V_c$, i.e. one such that $V \oplus iV \cong V_c$. The restriction of $B_c$ to $V$, call it $B$, is nondegenerate and it generates a Wick rule given by $B^{-1} = B_c^{-1}$. It is an element of $V_c \otimes_\bC V_c$ independent of the choice of the real subspace $V$. 

This is exactly what we want. Indeed, returning to the situation of interest, let $V_c = \Omega^{1,0}(\Sigma; \g_c)$. The holomorphic gauge propagator is $P_{hol} = B_c^{-1}$, the integral kernel of the inverse of $\bar\pd * \bar\pd$ on $\Omega^{1,0}(\Sigma; \g_c)$. The Feynman diagrammatic expansion generated by $\left<W_{f,\gm}\right>_{hol}$ using $P_{hol}$, in path integral notation, is given by
\begin{equation}
  \left<W_{f,\gm}\right>_{hol} = \frac{1}{Z}\int_V dA\, W_{f,\gm}(A)e^{-YM(A)} \label{eq:PI-axg}
\end{equation}
where $V \subset \A^{1,0}$ is any maximal totally real subspace. Indeed, this is simply the infinite-dimensional analog of what we discussed above.

It remains to discuss how (\ref{eq:PI-axg}) can be interpreted as a gauge-fixed path integral. The Yang-Mills action, originally defined on $\A$, extends holomorphically (i.e. complex-linearly) to the complexification $\A_c$ as discussed in Section \ref{sec:Setup}. As a consequence, this extended action is invariant under the action of the complex gauge group $\G_c$, which is the complexification of $\G$. Lemma \ref{Lemma:CGT} shows that
\begin{equation}
  \A_c = \G_c \cdot \Omega^{1,0}(\Sigma; \g_c) \label{eq:Gcslice}
\end{equation}
for $\Sigma = \R^2$ and for $\Sigma = S^2$ up to a negligible set of connections. An honest gauge-fixing condition is given by a (local) slice $\mc{S} \subset \A$, i.e. a submanifold transverse to the action of the gauge group $\G$. It yields for us the $\mc{S}$ gauge-fixed expectation
\begin{equation}
\left<W_{f,\gm}\right>_{\mc{S}} := \frac{1}{Z}\int_{\mc{S}}dA\, J_\mc{S}(A) W_{f,\gm}(A) e^{-YM(A)}. \label{eq:WLE-S} 
\end{equation}
Here, $J_{\mc{S}}(A)$ is the Faddeev-Popov determinant determined by the slice $\mc{S}$.

The slice $\mc{S}$ is a maximal (i.e. ``half-dimensional") totally real submanifold of $\mc{A}_c/\mc{G}_c$, since the latter is the complexification of $\mc{A}/\mc{G}$. On the other hand, $V$ is also a maximal totally real submanifold of $\mc{A}_c/\mc{G}_c$. In finite-dimensional integration theory, the integral of a holomorphic volume form is independent of the choice of half-dimensional real integration cycle (given the appropriate decay hypotheses so that integrals are convergent). Thus, changing the contour from $\mc{S}$ to $V$, from the holomorphicity of $YM(A)$ and $W_{f,\gm}(A)$ as functions of $A \in \mc{A}_c$, we formally obtain
\begin{equation}
  \left<W_{f,\gm}\right>_{hol} = \left<W_{f,\gm}\right>_{\mc{S}}. \label{eq:change-con}
\end{equation}
Indeed, we have the formal equality $\left<W_{f,\gm}\right>_{hol} = \left<W_{f,\gm}\right>_V$, with the latter defined as in (\ref{eq:WLE-S}). There is no Faddeev-Popov factor occuring in $\left<W_{f,\gm}\right>_V$, since it is a constant that factors out as is usual for axial-like gauges. Indeed, $J_V$ extends holomorphically to $\Omega^{1,0}(\Sigma;\g_c)$ and must be constant on this larger space, since the projection onto the complementary space $\Omega^{0,1}(\Sigma;\g_c)$ of the infinitesimal action of the complex gauge group is independent of $A \in \Omega^{1,0}(\Sigma;\g_c)$ (since $A$ has no $(0,1)$-part).

It is from the equality (\ref{eq:change-con}) that we can interpret holomorphic gauge as a gauge-fixing procedure. Alternatively, one can also use $\G_c$-invariance of the holomorphic extension of the path integral to deduce the following. Let $\mc{S}^{1,0} \subset \Omega^{1,0}(\Sigma; \g_c)$ be the unique submanifold such that $\G_c \cdot \mc{S}^{1,0} = \G_c \cdot \mc{S}$  up to a negligible set of connections, as is possible due to (\ref{eq:Gcslice}). Then via the action of $\G_c$, we can equate $\left<W_{f,\gm}\right>_{\mc{S}}$ with $\left<W_{f,\gm}\right>_{\mc{S}^{1,0}}$, with the latter defined as in (\ref{eq:WLE-S}) (and the Faddeev-Popov factor for $\mc{S}^{1,0}$ will be trivial as before, since it lies inside $\Omega^{1,0}(\Sigma;\g_c)$). One can then deform the contour $\mc{S}^{1,0}$ to the linear space $V$ while remaining inside $\Omega^{1,0}(\Sigma; \g_c)$ to obtain (\ref{eq:change-con}).

Thus, in the first approach, we (honestly) gauge-fix, complexify the quotient space, then deform the integration cycle within $\A_c/\G_c$ to a maximal totally real subspace $V$ of $\Omega^{1,0}(\Sigma;\g_c)$. In the second approach, we complexify the total space, complex gauge-fix to $\mc{S}^{1,0}$, and then deform within $\Omega^{1,0}(\Sigma;\g_c)$ to $V$. Because Yang-Mills theory is free in holomorphic gauge, we believe that the preceding heuristics may be convertible to rigorous mathematics that would, in particular, establish the perturbative area law to all orders in perturbation theory. We leave this to future investigation.

\end{document}